\documentclass[pra,twocolumn,showpacs,groupedaddress,superscriptaddress,aps,10pt]{revtex4-1}
\usepackage{bm,graphicx,amsmath}
\usepackage{placeins}
\usepackage{amssymb}
\usepackage{amsmath}
\usepackage{epsfig}
\usepackage{amssymb}
\usepackage{color}
\usepackage[colorlinks,linkcolor=blue,citecolor=blue]{hyperref}
\usepackage{subfigure}

\begin{document}

\title{Wave-vector and polarization dependence of conical refraction}
\date{\today}

\author{A. Turpin}
\affiliation{Departament de F\'isica, Universitat Aut\`onoma de Barcelona, Bellaterra, E-08193, Spain}
\author{Yu. V. Loiko}
\affiliation{Departament de F\'isica, Universitat Aut\`onoma de Barcelona, Bellaterra, E-08193, Spain}
\affiliation{Aston Institute of Photonic Technologies, School of Engineering \& Applied Science Aston University, Birmingham, B4 7ET, UK}
\author{T. K. Kalkandjiev}
\affiliation{Departament de F\'isica, Universitat Aut\`onoma de Barcelona, Bellaterra, E-08193, Spain}
\affiliation{Conerefringent Optics SL, Avda. Cubelles 28, Vilanova i la Geltr\'u, E-08800, Spain}
\author{H. Tomizawa}
\affiliation{JASRI/SPRING-8, 1-1-1, Kouto, Sayo, Hyogo, 679-5198, Japan}
\author{J. Mompart}
\affiliation{Departament de F\'isica, Universitat Aut\`onoma de Barcelona, Bellaterra, E-08193, Spain}

\begin{abstract} 
We experimentally address the wave-vector and polarization dependence of the internal conical refraction phenomenon by demonstrating that an input light beam of elliptical transverse profile refracts into two beams after passing along one of the optic axes of a biaxial crystal, \textit{i.e.} it exhibits double refraction instead of refracting conically. Such double refraction is investigated by the independent rotation of a linear polarizer and a cylindrical lens. Expressions to describe the position and the intensity pattern of the refracted beams are presented and applied to predict the intensity pattern for an axicon beam propagating along the optic axis of a biaxial crystal.  \\
\textbf{ocis}: (160.1190) Anisotropic optical materials; (260.1180) Crystal optics; (260.1140) Birefringence.
\end{abstract}

\date{\today}
\maketitle

\section{Introduction}

Light beams propagating in optically uniaxial crystals, with two principal dielectric indexes, exhibit birefringence or double refraction, \textit{i.e.} decomposition of an input beam into ordinary (o) and extraordinary (e) beams with orthogonal linear polarizations. The extraordinary beam is laterally shifted with respect to the ordinary one, being this shift equal to zero when the input beam propagates along the optic axis. The intensity distribution between the o- and e- beams, expressed by the Malus law, depends only on one parameter: the relative orientation of the crystal with respect to the polarization plane of the input beam. 

Biaxial crystals (BCs), with three different principal dielectric indexes and two optic axes, also exhibit double refraction. Nevertheless, propagation of light beams along the optic axis is substantially different from the uniaxial case. Hamilton predicted that, in this case, the beam should propagate conically inside the crystal and  emerge as a hollow cylinder yielding an annular transverse intensity pattern or light ring of internal conical refraction (CR) \cite{CRHistory}. Simultaneously to this prediction, Hamilton also demonstrated that an infinity of wave-vectors distributed on a hollow cone entering in the a biaxial crystal along the biradial propagate with common Poynting vector along the latter one, the so-called external conical refraction. Soon afterwards Hamilton's prediction, Lloyd reported the first experimental observation of both internal and external CR \cite{CRHistory} and  since then few experimental works have further studied CR experimentally (see \cite{Berry2} and references therein). Recently, internal CR has been applied to the laser technique \cite{CRlaser}, for optical micromanipulation \cite{Vortex} and for free space optical communications \cite{CRFSOC_OL}. Firsts theoretical efforts to describe precisely the internal conical refraction phenomenon were started by Portigal and Burstein \cite{port1969}. The complete calculation according to Fourier analysis was performed by Lalor \cite{lalor1972} and by Schell and Bloembergen \cite{schell1978} and later on rewritten by Belsky and Khapalyuk \cite{Belskii1,Belskii2}, Belsky and Stepanov \cite{Belsky} and Berry and Jeffrey \cite{Berry2}. From the experimental point of view, the full validation of the internal CR was given by F\`eve and coworkers \cite{Feve1994}. However, a simple analytical formulation capable of predicting the transverse intensity pattern at the Lloyd (focal) plane after a light beam propagates along or near the optic axis of a biaxial crystal (or a cascade of BCs \cite{Berry3, abd2011}) is still lacking. 

In this article, we experimentally address this last need by analyzing the wave-vector and polarization dependence of the internal CR phenomenon. In the following we will refer to internal CR simply as CR. Our experimental scheme is based on the propagation of elliptical beams (EBs) along the optic axis of a BC. EBs do not possesses continuous cylindrical symmetry, at variance with Gaussian beams typically used in conical refraction experiments. In Section 2, we demonstrate that EBs do not produce the otherwise expected transverse annular intensity pattern when they propagate through the BC, but split into two refracted beams with orthogonal linear polarizations. Simple phenomenological expressions for the position and the intensity of the refracted beams are derived in Section 3. Worth noticing, we show that the position of the refracted beams is defined not only by the crystal orientation, but also by the input beam properties, in particular, by the orientation of the preselected plane of input wave-vectors. Additionally, we demonstrate that the intensity distribution between the two refracted beams can be well described with a single angular parameter. When the biaxial crystal is rotated, the intensity distribution between the two refracted beams follows the transformation law recently pointed out by Loiko~\textit{et al.}~\cite{Loiko} for CR filtered beams, that differs from the well known Malus law for double refraction in uniaxial crystals. To verify the usefulness of the phenomenological transformation rules, we apply them in Section 4 to calculate the resulting intensity pattern for an incident axicon beam and confirm our approach with experimental results. Section 5 summarizes the main results and conclusions of the work.

\section{Conical refraction of spatially anisotropic beams}

\begin{figure}[h!]
\centering\includegraphics[width=1.0 \columnwidth]{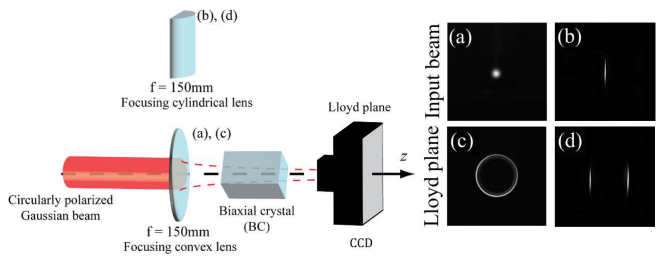}
\caption{Experimental set-up of light refraction along the optic axis of a biaxial crystal (BC): the input beam is focused by a lens and passes through the BC. The refraction pattern is obtained at the Lloyd plane of the system, which coincides with the focal plane. (a) An input Gaussian beam focused with a spherical lens (focal length of $150\,\rm{mm}$) yields the well-known ring of CR (c) at the Lloyd (focal) plane of the system. (b) An elliptical beam (EB) beam, obtained  with a cylindrical lens (focal length of $150\,\rm{mm}$) that focuses in the horizontal direction (see Fig.\ref{fig2} for details), yields the the double refraction pattern shown in figure (d).}
\label{fig1}
\end{figure}

Conical refraction has been reported only for input beams with intensity pattern possessing continuous cylindrical (rotation) symmetry around the propagation axis, see Fig.~\ref{fig1}(a) and Fig.~\ref{fig1}(c). This has been motivated by Hamilton's prediction of the CR within the ray optics description, where rays are always associated with cylindrically symmetric collimated beams. Real beams of finite size can be modeled as a bundle of rays with propagation directions  isotropically distributed around the beam axis. On the one hand, according to the modified Hamilton theory \cite{Berry2}, CR can be interpreted as a wave-vector and polarization dependent type of double refraction, when each ray of the input bundle refracts into two rays at the entrance crystal surface and all refracted rays form a cone. In this process, the two refracted rays go to opposite points on the CR ring. On the other hand, it has been also demonstrated that in CR the transverse wave-vector components are conserved \cite{Berry2, Berry3}. This means that all input rays (plane waves) with wave-vectors confined in a certain plane defined by an azimuthal angle $\phi$ refract to a plane with the same azimuthal angle at the Lloyd plane, \textit{i.e.} into two opposite points on the CR ring. 

Therefore, double refraction along the optic axis of a BC and, in particular, the wave-vector and polarization dependence of the CR, can be studied with linearly polarized beams composed from plane waves whose wave-vectors are confined in a certain plane. Such beams are known as cylindrical beams and elliptical beams (EBs) provide their finite size approximation. The latter ones can be obtained from collimated Gaussian beams focused by a cylindrical lens. Indeed, we have observed that an EB passing through the BC does not generate the CR ring, but splits into two beams (in case of circularly polarized input beams) oppositely placed along the otherwise expected CR ring and tangent to it, see Fig.~\ref{fig1}(d). The form and orientation of the refracted beams resemble the transverse intensity pattern and orientation of the input EB. Below, using EBs we will obtain phenomenological laws that describe double refraction along the optic axis in biaxial crystals.

\section{Transformation rules of conical refraction}

\begin{figure}[h!]
\centering\includegraphics[width=1.0 \columnwidth]{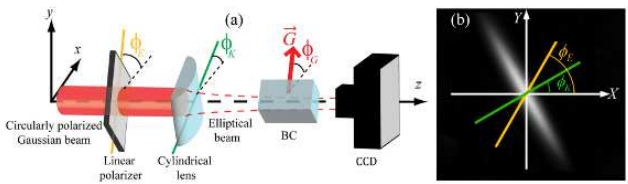}
\caption{(a) An elliptical beam with linear polarization is generated when a circularly polarized collimated Gaussian beam passes through the linear polarizer (with polarization plane given by azimuthal angle $\phi_{E}$) and is focused by a cylindrical lens (focal length of $150\,\rm{mm}$), which determines the wave-vectors plane (given by azimuthal angle $\phi_{K}$). The resulting patterns are captured by a CCD camera at the Lloyd plane behind the BC. The orientation of the crystal is characterized by the orientation of the plane of its optic axes (given by azimuthal angle $\phi_{G}$). (b) Elliptical beam at the focal plane of the lens when the BC is removed.
The beam is parameterized by the azimuthal angles $\phi_{E}$ and $\phi_{K}$ related to the polarization and wave-vector planes, respectively. All angles are measured from the horizontal $x$-axis of the laboratory system of coordinates.}
\label{fig2}
\end{figure}

In this section, we will investigate the position and the relative intensity of the two refracted beams produced when an input EB propagates along the optic axis of a BC. The experimental set-up shown in Fig.~\ref{fig2}(a) is used. The initial circularly polarized Gaussian beam with waist radius of $w = 1\,\rm{mm}$ is obtained from a $640\,{\rm nm}$ diode laser coupled to a monomode fiber. Then a linear polarizer is introduced to fix the polarization plane in a well defined direction. The resulting linearly polarized beam is focused by a cylindrical lens of $150\,\rm{mm}$ focal length with its flat face oriented strictly perpendicular to the beam propagation direction. The cylindrical lens only focuses the Gaussian beam in one direction, so that it transforms its transverse circular shape to an elliptical one with a ratio 3/100 of the semi-axes of the ellipse. As a consequence, the divergence of the generated EB is different along the focused and unfocused directions ($w_{f} = 30\,{\rm \mu m}$, $\theta_{f} = 6.8\,{\rm mrad}$; $w_{uf} = 1000\,{\rm \mu m}$, $\theta_{uf} = 0.2\,{\rm mrad}$, where \textit{f} and \textit{uf} subscripts refer to \textit{focused} and \textit{unfocused} directions, respectively).
The EB is characterized by its polarization plane, represented by the azimuthal angle $\phi_{E}$, and by its plane of wave-vectors (or $K$-plane), represented by azimuthal angle $\phi_{K}$; see Fig.~\ref{fig2}(b). Different EBs are obtained by rotating either the cylindrical lens or the linear polarizer. The BC is $28\,\rm{mm}$ long and it was cut from a monoclinic centrosymmetric KGd(WO$_4$)$_2$ crystal. Its polished entrance surfaces (cross-section $6\times 4 \,\rm{mm}^2$, parallelism 10 arc seconds) are perpendicular to one of the optic axes (misalignment angle $<1.5\,\rm{mrad}$) and positioned strictly perpendicular to the beam propagation direction, so that the incoming beam to the BC passes along its optic axis. Both the orientation of the optic axis of the BC and the cylindrical lens are well controlled in the $\theta$ and $\varphi$ directions in 3D spherical coordinates by a micrometer positioning system. The resulting pattern is captured by a CCD camera at the Lloyd (focal) plane behind the BC \cite{Todor}.

\subsection{Position of the refracted beams}

First, we briefly outline the CR annular pattern observed at the Lloyd plane with an input Gaussian beam. While the complete CR ring is observed for circular polarization of the input Gaussian beam, as it is shown in Fig.~\ref{fig1}(d), a crescent ring with one point of the ring being of zero intensity appears for linear polarization. In both cases, the center of the CR ring is laterally shifted with respect to the incident beam. This shift can be represented by a vector $\mathbf{G}= R_{0} \left( \cos \phi_{G}, \sin \phi_{G}\right)$ that belongs to the plane of the crystal optic axes and is perpendicular to them \cite{Todor}. Its modulus is equal to the ring radius $|\mathbf{G}| \equiv R_0$ (in our set-up $ R_0= 476\,\mu\rm{m}$), given by a product of the length and the conical refraction semiangle of the BC ($17\,\rm{mrad}$ for KGd(WO$_4$)$_2$) \cite{Todor}.

\begin{figure}[h!]
\centering\includegraphics[width=1.0 \columnwidth]{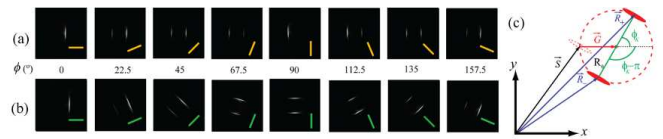}
\caption{Transverse intensity patterns obtained after rotating (a) the polarizer, \textit{i.e.} varying $\phi_{E}$, or (b) cylindrical lens, \textit{i.e.} varying $\phi_{K}$. $\phi$ means $\phi_{E}$ in (a) and $\phi_{K}$ in (b) and it is varied in the range $\left[ 0^{\circ}, 157.5^{\circ} \right]$ with steps of $22.5^{\circ}$, while $\phi_{G}=0^{\circ}$. Yellow (a) and green (b) lines at the bottom right corner indicate the polarization $E$-plane (a) and the wave-vector $K$-plane (b), respectively. (c) Splitting of an input EB beam (dashed ellipse), where $\mathbf{R}_{\pm}$ denote the position of the two output refracted beams at the Lloyd plane. Dashed ring in (c) denotes the otherwise expected CR in case of input beam of Gaussian profile.}
\label{fig3}
\end{figure}

Now we will experimentally deduce the lateral shift of the refracted beams for EBs propagated through a BC.
In Fig.~\ref{fig3} two series of images present the transverse intensity pattern at the Lloyd plane recorded varying either $\phi_{E}$ ($\phi_{K}=0^{\circ}$) (a) or $\phi_{K}$ ($\phi_{E}=0^{\circ}$) (b) from $0$ to $157.5^{\circ}$ in $22.5^{\circ}$ intervals, when the crystal orientation remains fixed at $\phi_{G}=0^{\circ}$. The geometric center of the refracted beams coincides with the center of the otherwise expected CR ring, since it is shifted from the position of the initial EB by the vector \textbf{G}. This shift is shown schematically in Fig.~\ref{fig3}(c), and it is subtracted in Fig.~\ref{fig3}(a) and Fig.~\ref{fig3}(b). 

From Fig.~\ref{fig3} it is clear that in a local frame with origin at the ring center, the position of the two refracted beams does not depend on the input beam polarization, $\phi_{E}$, but rotates linearly with $\phi_{K}$. Moreover, the azimuthal angles $\phi_{\pm}$ of the refracted beams are defined by the wave-vector plane, $\phi_{K}$, of the input EB, namely, 
\begin{equation}
\phi_+ = \phi_{K},~ \phi_- = \phi_{K} + \pi. 
\label{eqphipm}
\end{equation}
Since $\phi_{K}$ and $\phi_{K} + \pi$ describe the same wave-vector plane, the latter expressions mean conservation of the wave-vectors $K$-planes. Rotation of the crystal, \textit{i.e.} change of $\phi_G$, does not affect the angles $\phi_{\pm}$, but it affects the position of the center of the refracted beams and redistributes the intensity between the refracted beams as it will be shown in the next subsection.
Summarizing, in the $xy$ laboratory coordinates, the positions of the refracted beams $\mathbf{R}_{\pm}$ at the Lloyd plane can be written as follows:

\begin{equation}
\mathbf{R}_{\pm}(\phi_{K})=\mathbf{S}(\phi_{K}) + \mathbf{G} \pm R_0 \mathbf{u}(\phi_{K}),
\label{eq1}
\end{equation}
where $\mathbf{u}(\phi_K) \equiv \left(\cos\left(\phi_K \right), \sin\left(\phi_K \right)\right)$. $\mathbf{S}$ denotes the position at the Lloyd plane where the initial EB would be focused in the absence of the BC, see Fig.~\ref{fig3}(c). In other words, the two refracted beams are located at diagonal positions of the otherwise expected CR ring. EBs obtained from the same Gaussian beam have the same initial position $\mathbf{S}$, \textit{i.e.} $\mathbf{S}$ does not depend on $\phi_{E,K,G}$ in this case. As a final comment, Eqs.~(\ref{eq1}) generalize the geometrical approach \cite{Berry2}.

\subsection{Relative intensity distribution of the refracted beams}

\begin{figure}[h!]
\centering\includegraphics[width=1.0 \columnwidth]{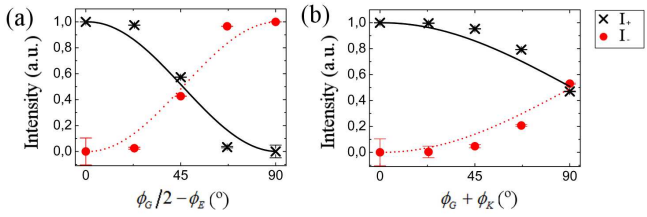}
\caption{Normalized intensities of the two refracted beams after the splitting of the elliptical beam as a function of the variation of the parameters (a) $\phi_{G} /2 - \phi_{E}$ ($\phi_{K} = 0^{\circ}$) or (b) $\phi_{G}+\phi_{K}$ ($\phi_{E} = 0^{\circ}$). Black solid ($I_{+}$) and red dashed ($I_{-}$) curves show the analytical fitting given by Eqs.(\ref{eqintensity}), while symbols represent the corresponding experimental data. The error in angle measurements is $\pm 0.5^\circ$.}
\label{fig4}
\end{figure}

Now we report how the intensity of the input EB is distributed between the two refracted beams. In the experiments, three parameters can be varied independently: $\phi_{G}$, $\phi_{E}$ and $\phi_{K}$, associated to the crystal orientation represented by $\mathbf{G}$ and to the polarization and wave-vectors planes of the incident EBs. Below we show that only one combination of these angles governs the relative intensity distribution between the two refracted beams. With this purpose, we have repeated the experiments shown in Fig.~\ref{fig3}(a) and Fig.~\ref{fig3}(b) for different orientations of the BC. Symbols (black crosses and red circles) in Fig.~\ref{fig4}(a) Fig.~\ref{fig4}(b) show the corresponding experimental results for the intensities $I_{\pm}$ of the two refracted beams normalized with respect to the intensity of the incident beam. Black solid and red dashed curves represent their analytical fittings given by the following expressions:
\begin{eqnarray}
I_{+} \{\phi_{E},\phi_{G},\phi_{K}=0 \} = I_0 \cos^2 \left( \frac{\phi_{G}}{2} - \phi_{E} \right);\\
I_{-} \{\phi_{E},\phi_{G},\phi_{K}=0 \} = I_0 \sin^2 \left( \frac{\phi_{G}}{2} - \phi_{E} \right), 
\label{eqexp1}  \\
I_{+} \{\phi_{E}=0,\phi_{G},\phi_{K} \} = I_0 \cos^2 \left( \frac{\phi_{G}}{2} + \frac{\phi_{K}}{2} \right);\\
I_{-} \{\phi_{E}=0,\phi_{G},\phi_{K} \} = I_0 \sin^2 \left( \frac{\phi_{G}}{2} + \frac{\phi_{K}}{2} \right).
\label{eqexp2}
\end{eqnarray}
$I_{+}$ and $I_{-}$ are the intensities of the beams refracted at angles $\phi_{+}$ and $\phi_{-}$ and located diagonally at the both ends of the CR ring at positions $R_+$ and $R_-$ respectively, following Eq.(\ref{eq1}).

Eqs.~(\ref{eqexp1}) and (\ref{eqexp2}) can be rewritten in a unified simple form, giving the next formula for the relative intensity distribution between the two refracted beams with only one governing parameter $\omega$:
\begin{equation}
I_{+} = I_{0} \cos^2 \left( \frac{\omega}{2} \right),
\quad 
I_{-} = I_{0} \sin^2 \left( \frac{\omega}{2} \right),
\quad \\
\omega = \phi_{G} - \phi_{\chi},
\quad
\phi_{\chi} = 2 \phi_{E} - \phi_{K}.
\label{eqintensity}
\end{equation}
From Eq.~(\ref{eqintensity}) it follows that, with respect to the relative energy distribution the only significant parameters of the input EB is $\phi_{\chi}$. Therefore, the intensity splitting under CR can be expressed in terms of the difference between the parameter $\phi_{\chi}$  and the orientation of the BC, given by $\phi_{G}$. We would like to highlight here that the expression $\phi_{\chi}=2 \phi_{E} - \phi_{K}$ that we have obtained experimentally, corrects the theoretical result derived in \cite{Berry3}, where a plus sign was considered instead of a minus one.
Eqs.~(\ref{eq1}) and (\ref{eqintensity}) describe also recent experimental results on the propagation of CR filtered beams \cite{Loiko} along the optic axis of a BC. In that case, $\phi_{\chi}=\phi_{\mathbf{G}}$, where $\phi_{\mathbf{G}}$ denotes the orientation of an initial biaxial crystal that produces a CR ring with lateral shift given by vector $\mathbf{G}_0$. Then, a pinhole placed at angle $\phi$ on the CR ring produces a CR filtered beam with polarization $\phi_{E} = \frac{\phi}{2}$ and plane of wave-vectors $\phi_{K} = \phi $ that also splits into two beams after passing along the optic axis of a biaxial crystal . 

Eqs.~(\ref{eq1}) and (\ref{eqintensity}) constitute the Transformation Rules of Conical Refraction. For different polarization of the input beams they allow explaining the ratio of intensities for any pair of diagonally opposite points of the CR ring. These results corroborate with experimental observations that the complete CR ring appears only for beams with azimuthally continuous symmetric distribution of wave-vectors. As a proof of usefulness of the formalism introduced in this work, in the next section we apply the derived transformation rules to an axicon input beam.

\section{Application of the transformation rules of CR to an axicon beam}

The experimental set-up is sketched in Fig.~\ref{fig5}. A linearly polarized conical beam is prepared when an initial circularly polarized Gaussian beam passes through a linear polarizer and an axicon lens. The axicon beam is focused by a spherical lens into the BC. The resulting pattern is captured with a CCD camera at the focal plane. As in previous experiments, the orientation of the optic axis of the BC, the spherical lens and the axicon lens are well controlled in the $\theta$ and $\varphi$ directions in 3D spherical coordinates by a micrometer positioning system. Each infinitesimally thin azimuthal sector of the axicon lens characterized by azimuthal angle $\phi$ forms a thin prism that produces a wave with particular wave-vector whose transverse projection comprise an angle $\phi_{k}=\phi$. The axicon lens generates, therefore, a continuous collection of beams with $\phi\in [ 0, 2 \pi )$. After the axicon, the refracted beams, following  Snell law, have the same inclination angle $\theta_0$ with respect to the $z$-axis. At the focal plane of the lens they form a ring such that each point can be characterized by wave-vector plane, $\phi_{k}$, and polarization plane, $\phi_{E}$.
In other words, each point of the axicon ring is an EB. In this case all these EB have their polarization plane fixed at $\phi_E$ and their wave-vector plane $\phi$ is varying continuously along the ring as shown in Fig.~\ref{fig6}(a). Behind the BC, the refraction pattern can be calculated by applying Eq.~(\ref{eq1}) to every point of the input axicon annular beam taking into account the initial positions as given by  $\mathbf{S} \left(\phi\right)=R_{ax} \left(\cos{\phi}, \sin{\phi} \right)$ (where $R_{ax}$ is the radius of the axicon light ring). Therefore, from Eq.~(\ref{eq1}) one can obtain the refracted pattern for an axicon beam:
\begin{figure}[]
\centering\includegraphics[width=1.0 \columnwidth]{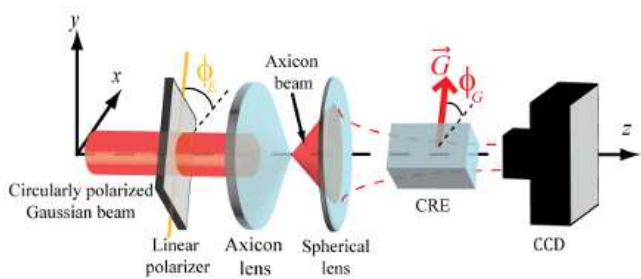}
\caption{Experimental set-up for axicon beam propagation	along the optic axis of a biaxial crystal. The axicon lens (apex angle of 179.5$^{\circ}$) generates a conical beam from an input linearly polarized Gaussian beam which is then focused by a spherical lens (focal length of $150\,\rm{mm}$) along the optic axis of the BC.}
\label{fig5}
\end{figure}
\begin{figure}[h!]
\centering\includegraphics[width=1.0 \columnwidth]{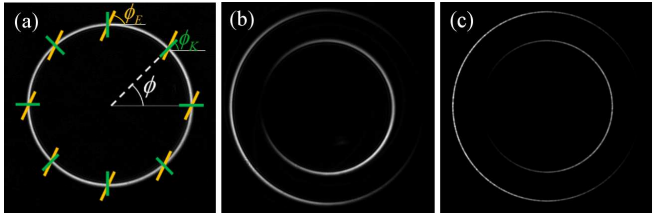}
\caption{Refraction of linearly polarized axicon annular beam along optic axis of a biaxial crystal. (a) Schematic representation of linearly polarized axicon annular beam with each point characterized by the azimuthal angles $\phi_{k}=\phi$ (short green  lines) and all of them have the same polarization plane $\phi_{E}$ (short orange lines). (b) Experimentally observed transverse intensity pattern at the Lloyd plane consisting of two concentric rings when a linearly polarized axicon beam with $\phi_E=0$ propagates along the optic axis of the biaxial crystal with $\phi_{G}=0$. (c) Corresponding theoretical simulation from Eqs.~(\ref{eqaxicon}) and (\ref{eqintaxicon}).}
\label{fig6}
\end{figure}  
\begin{equation}
\mathbf{R}_{\pm}(\phi) = \mathbf{G}+\left(  \left| R_{ax} \right| \pm \left| R_{0} \right| \right)\mathbf{u}(\phi).
\label{eqaxicon}
\end{equation}
These expressions, with $\phi$ scanned from $0$ to $2\pi$, parameterize two concentric rings with radii $R_{ax} \pm R_0$ laterally shifted by $\mathbf{G}$ relatively to the axicon ring axis. The intensity distribution is calculated from Eq.~(\ref{eqintensity}). All points of the incident axicon beam have the same intensity, which is distributed between the two refracted rings as follows:  
\begin{equation}
I_+ = I_0 \frac{R_0}{R_+}  \cos^2 \left( \frac{\phi-\phi_0}{2} \right), \quad 
I_- = I_0 \frac{R_0}{R_-}  \sin^2 \left( \frac{\phi-\phi_0}{2} \right),
\label{eqintaxicon}
\end{equation}
where we have taken $\omega = \phi - \phi_0$ being $\phi_0 \equiv 2\phi_E - \phi_G$ a constant parameter. In addition, since both rings have different radii, normalization factors $R_{0}/R_{\pm}$ have been introduced to $I_{\pm}$ to assure energy conservation. Fig.~\ref{fig6}(a) shows the intensity pattern of the axicon beam with polarization and $K$-plane distribution. The experimental refraction pattern behind the BC is shown in Fig.~\ref{fig6}(b). The pattern is formed by two concentric rings oppositely polarized, with polarization distribution analogous to that one obtained in a cascaded CR configuration \cite{Todor}. 
Polarization of the ring points with minimum (zero) and maximum intensities are orthogonal and parallel respectively to that of the input axicon beam. Fig.~\ref{fig6}(c) presents the theoretical prediction of the light refraction pattern  from Eqs.(\ref{eqaxicon}) and (\ref{eqintaxicon}). Comparison between Fig.~\ref{fig6}(b) and Fig.~\ref{fig6}(c) shows that the theoretical prediction and the experiment (see Fig.~\ref{fig6}(b)) are in complete agreement. 

\section{Conclusions}

It is well known that an input Gaussian beam produces a light ring after propagating along the optic axis of a biaxial crystal. In this paper we have experimentally addressed the wave-vector and polarization dependence of the internal conical refraction phenomenon in biaxial crystals. We have experimentally proved that this phenomenon is only one particular case of refraction associated to beams possessing continuous axial cylindrical symmetry. In fact, we have reported that elliptical beams -- obtained by focusing a Gaussian beam with a cylindrical lens -- do not generate the complete characteristic light ring of conical refraction when they propagate along an optic axis of a biaxial crystal but split, in general, into two beams. We have phenomenologically obtained expressions for the positions and relative intensities of the resulting refracted beams. These expressions play an analogous role as the Malus law but for biaxial crystals and can be used to predict the transverse intensity pattern of much more involved incident beams as it has been demonstrated here for an input axicon beam. 

\section{Acknowledgement}
The authors gratefully acknowledge financial support through Spanish MICINN contracts FIS2010-10004-E and FIS2011-23719, and the Catalan Government contract SGR2009-00347. A. T. acknowledges financial support through grant AP2010-2310 from the MICINN. Dr. Jos\'e Luis Mart\'inez is also acknowledged for his helpful comments.


\begin{thebibliography}{99}

\bibitem{CRHistory} J. G. O'Hara, ``The prediction and discovery of conical refraction by William Rowan Hamilton and Humphrey Lloyd,'' Proc. R. Ir. Acad. \textbf{82} (2), 231--257 (1982).

\bibitem{Berry2} M. V. Berry and M. R. Jeffrey, ``Conical diffraction: Hamiltonӳ diabolical point at the heart of crystal optics,'' Prog. Opt. \textbf{50}, 13--50 (2007).

\bibitem{CRlaser} A. Abdolvand, K.G. Wilcox, T. K. Kalkandjiev, and Edik U. Rafailov, ``Conical refraction $Nd:KGd(WO_4)_2$ laser,'' Optics Express \textbf{18}, 2753--2759 (2010).

\bibitem{Vortex} D. P. O'Dwyer, K. E. Ballantine, C. F. Phelan, J. G. Lunney, and J. F. Donegan, ``Optical trapping using cascade conical refraction of light,'' Optics Express \textbf{20}, 21119--21125 (2012).

\bibitem{CRFSOC_OL} A. Turpin, Yu. V. Loiko, T. K. Kalkandjiev and J. Mompart, ``Free-space optical polarization demultiplexing and multiplexing by means of conical refraction,'' Opt. Lett. \textbf{37}, 4197--4199 (2012).

\bibitem{port1969} D. L. Portigal and E. Burstein, ``Internal Conical Refraction,'' J. Opt. Soc. Am.  \textbf{59}, 1567--1573 (1969).

\bibitem{lalor1972} E. Lalor, ``An Analytical Approach to the Theory of Internal Conical Refraction,'' J. Math. Phys. \textbf{13}, 449--454 (1972).

\bibitem{schell1978} A. J. Schell and N. Bloembergen, ``Laser studies of internal conical diffraction. I. Quantitative
comparison of experimental and theoretical conical intensity distribution in aragonite,'' J. Opt. Soc. Am.  \textbf{68}, 1093--1098 (1978).

\bibitem{Belskii1} A. M. Belskii and A. P. Khapalyuk, ``Internal conical refraction of bounded light beams in
biaxial crystals,'' Opt. Spectrosc. \textbf{44}, 436֭439 (1978). 

\bibitem{Belskii2} A. M. Belskii and A. P. Khapalyuk, ``Propagation of confined light beams along the beam axes (axes of single ray velocity) of biaxial crystals,'' Opt. Spectrosc. \textbf{44}, 312--315 (1978).

\bibitem{Belsky} A. M. Belsky and M.A. Stepanov, ``Internal conical refraction of light beams in biaxial gyrotropic crystals,'' Optics Commun. \textbf{204}, 1֭6 (2002).

\bibitem{Feve1994} J. P. F\`eve, B. Boulanger and G. Marnier, ``Experimental study of internal and external conical refractions
in KTP,'' Optics Commun. \textbf{105}, 243--252 (1994).

\bibitem{Berry3} M.V. Berry, ``Conical diffraction from an N-crystal cascade,''
\textit{J. of Optics}, \textbf{12}, 075704 (2010).

\bibitem{abd2011} A. Abdolvand, ``Conical diffraction from a multi-crystal cascade: experimental observations,'' \textit{Appl. Phys B} \textbf{103}, 281--283 (2011).

\bibitem{Loiko} Y. Loiko, M. A. Bursukova, T. K. Kalkanjiev, E. U. Rafailov, and J. Mompart, ``Fermionic transformation rules for spatially filtered light beams in conical refraction,'' in \textit{Complex Light and Optical Forces V}, D. L. Andrews, E. J. Galvez, J. Gl\"uckstad, eds., Proc. SPIE \textbf{7950}, 79500D--79500D-9 (2011).

\bibitem{Todor} T. K. Kalkandjiev and M. A. Bursukova, ``Conical refraction: an experimental introduction,'' in \textit{Photon Management III}, J. T. Sheridan, F. Wyrowski, eds., Proc. SPIE \textbf{6994}, 69940B--69940B-10 (2008).

\end{thebibliography}
\end{document}